\documentclass[twocolumn,twocolappendix,tighten,resetfootnote]{aastex701}

\usepackage{amsmath}
\usepackage{graphicx}
\usepackage{booktabs}
\usepackage{multirow}
\usepackage{xcolor}
\usepackage{soul}
\setstcolor{red}

\shorttitle{Proxima Centauri Angular Diameter}
\shortauthors{Anugu et al.}

\begin{document}

\title{Proxima Centauri's Angular Diameter Revisited}

\correspondingauthor{Narsireddy Anugu}
\email{nanugu@gsu.edu}

\author[0000-0000-0000-0000]{Narsireddy Anugu}
\thanks{NA and AG contributed equally to this work.}
\affiliation{The CHARA Array of Georgia State University, Mount Wilson Observatory, Mount Wilson, CA 91023, USA}
\email{nanugu@gsu.edu}

\author[0000-0001-7853-4094]{Alexandre Gallenne}
\affiliation{Instituto de Alta Investigaci\'on, Universidad de Tarapac\'a, Casilla 7D, Arica, Chile}
\email{agallenne@academicos.uta.cl}

\author[0000-0003-0626-1749]{Pierre Kervella}
\affiliation{LIRA, Observatoire de Paris, Universit\'e PSL, Sorbonne Universit\'e, Universit\'e Paris Cit\'e, CY Cergy Paris Universit\'e, CNRS, 5 place Jules Janssen, 92195 Meudon, France}
\affiliation{French-Chilean Laboratory for Astronomy, IRL 3386, CNRS and U. de Chile, Casilla 36-D, Santiago, Chile}
\email{pierre.kervella@obspm.fr}

\author[0000-0003-3486-853X]{Adam D. Rains}
\affiliation{Research School of Astronomy and Astrophysics, Australian National University, Canberra, ACT 2611, Australia}
\affiliation{Instituto de Astrofísica, Pontificia Universidad Católica de Chile, Av. Vicuña Mackenna 4860, 782-0436 Macul, Santiago, Chile}
\email{adam.rains@uc.cl}

\begin{abstract}
Proxima Centauri is the nearest known exoplanet-hosting star and a benchmark for low-mass stellar models. Published diameters differ by several percent, affecting its inferred radius and temperature. Model-dependent near-infrared limb darkening introduces an additional 2--5\% correction to the inferred radius. We re-reduced two previously published VLTI/PIONIER interferometric epochs from 2014 and combined them with three previously unpublished archival epochs from 2017 and 2019. We fit all five data sets jointly with uniform-disk and atmosphere-grid-based center-to-limb models. The joint fit gives a uniform-disk angular diameter, $\theta_{\rm UD}=1.117\pm0.024$~mas. Using the PHOENIX model and adopted bolometric flux, we obtain a PHOENIX limb-darkened angular diameter, $\theta_{\rm LD,PHX}=1.154\pm0.026$~mas, a stellar radius, $R_\star=0.162\pm0.004\,R_\odot$, and a Stefan--Boltzmann effective temperature, $T_{\rm eff,SB}=2833\pm49$~K. No significant close companion is detected across the five epochs over the searched separation and contrast range.
\end{abstract}


\section{Introduction}

Proxima Centauri (GJ~551; hereafter Proxima) is the nearest known exoplanet host and a fully convective M-dwarf star that provides a stringent test of low-mass stellar models. Its angular diameter and distance determine the stellar radius, while its limb-darkened diameter and bolometric flux determine the effective temperature. Accurate values of these quantities are important for characterizing Proxima and interpreting the properties and potential habitability of Proxima~b.

For more than a century, Proxima's angular diameter has challenged observers: from the early estimate of \citet{Plaskett1922} to modern long-baseline interferometry, its small angular size has made a precise measurement difficult.
Modern Very Large Telescope Interferometer (VLTI) observations with VINCI and AMBER yielded limb-darkened diameters of $1.02\pm0.08$~mas and $1.011\pm0.052$~mas, respectively \citep{Segransan2003,Demory2009}.
The most precise published result was obtained with VLTI/PIONIER by \citet{Rabus2019}, who measured a uniform-disk diameter $\theta_{\rm UD}=1.066\pm0.007$~mas and, after applying a model-dependent limb-darkening correction, $\theta_{\rm LD}=1.103\pm0.007$~mas from 2014 observations. These measurements made Proxima a useful, well-characterized benchmark for models of low-mass stars.

An interferometer infers angular size from the squared visibility $V^2$, which quantifies fringe contrast as a function of projected baseline. For a uniform disk, $V^2$ decreases with baseline and reaches zero at the first visibility null once the star is sufficiently resolved. Although long-baseline $H$-band PIONIER observations \citep{LeBouquin2011} provide the required angular resolution, Proxima is only partially resolved, i.e., all $V^2$ data points lie before this first null. Its diameter must therefore be inferred from small changes in the calibrated squared visibility rather than from a directly measured null, making the result sensitive to instrumental and calibration systematics. \citet{Lachaume2019} demonstrated the importance of treating correlated errors for such under-resolved stars.

Previous PIONIER studies obtained consistent diameters \citep{Rabus2019,Lachaume2019}, but their measurements were based on only two observing epochs, both obtained in 2014. Because Proxima is only partially resolved, small calibration or reduction effects can influence the inferred angular scale. To test whether that scale is robust to these effects, we re-reduce the raw ESO archive data for both 2014 epochs and three unpublished epochs from 2017 and 2019 with one procedure, then fit all five epochs jointly. The joint fit tests whether the angular scale is repeatable across observing conditions and calibrator sets while reducing the influence of any single epoch. We use it to determine the $H$-band diameter, quantify the model-dependent limb-darkening conversion, infer the radius and effective temperature, and search the same data for a close companion.

The paper is organized as follows. Section~\ref{sec:observations} describes the observations and data selection, the reduction and calibration procedure. The methods used for visibility modeling, uncertainty estimation, and the companion search are detailed in Section~\ref{sec:data_analysis}. Section~\ref{sec:results} presents the angular-diameter measurements and the resulting radius and effective temperature, together with the companion-detection limits. Section~\ref{sec:discussion} compares our results with previous measurements and discusses the principal observational and model-dependent limitations. Section~\ref{sec:conclusions} summarizes our conclusions.

\begin{table*}[t]
\centering
\caption{VLTI/PIONIER Proxima observing blocks and calibration information.}
\label{tab:calibrators}
\scriptsize
\begin{tabular}{llll}
\toprule
Epoch & ESO programme & PI & \shortstack[l]{Calibrators ($\theta_{\rm UD}$, mas)} \\
\midrule
2014 Mar 21 & 092.D-0647 & Rabus & HD~119073 ($0.399\pm0.009$) \\
2014 Jul 11 \& 12 & 093.D-0471 & Rabus & HD~128398 ($0.351\pm0.008$), HD~128917 ($0.379\pm0.009$) \\
2017 Jul 27 & 299.C-5039(B) & Kervella & HD~128713 ($0.979\pm0.094$), HD~125640 ($0.455\pm0.012$)  \\
2019 Jan 16 & 0102.D-0562(A) & Rains & HR~5349 ($0.397\pm0.031$), HD~125288 ($0.405\pm0.040$) \\
2019 Feb 22 \& 23 & 0102.D-0562(A) & Rains & HR~5349 ($0.397\pm0.031$), HD~125288 ($0.405\pm0.040$)  \\
\bottomrule
\end{tabular}
\vspace{2pt}
\begin{minipage}{0.95\textwidth}
\footnotesize\textit{Notes.} The 2014 epochs used the A1--G1--J3--K0 Auxiliary Telescope configuration; the 2017 and 2019 February epochs used A0--G1--J2--J3, and the 2019 January epoch used A0--G2--J2--J3. Listed values are the adopted uniform-disk calibrator diameters and $1\sigma$ errors used for the calibrated OIFITS products. The calibrators were selected with JMMC SearchCal and are drawn from stellar-diameter catalogues described by \citet{Chelli2016,Bourges2017}.
\end{minipage}
\end{table*}

\begin{table*}[t]
\centering
\caption{Joint five-epoch angular-diameter fits to the VLTI/PIONIER data.}
\label{tab:diameter-fits}
\scriptsize
\setlength{\tabcolsep}{3pt}
\begin{tabular}{lccc}
\toprule
\shortstack[l]{Model and\\diameter definition} &
\shortstack{$\theta_{\rm fit}$\\Fitted diameter (mas)} &
\shortstack{$\alpha_{\rm prior}$\\Prior $\mu\pm\sigma$} &
\shortstack{$\alpha_{\rm post}$\\Posterior 68\% interval} \\
\midrule
Uniform disk ($\theta_{\rm UD}$) &
$1.117\pm0.024$ & --- & --- \\
Power law ($\theta_{\rm PL}$): Kurucz prior &
$1.136\pm0.027$ & $0.124825\pm0.000084$ & $0.124825\pm0.000068$ \\
Power law ($\theta_{\rm PL}$): Stagger prior &
$1.151\pm0.028$ & $0.241915\pm0.008619$ & $0.241980\pm0.006971$ \\
Power law ($\theta_{\rm PL}$): MPS1 prior &
$1.146\pm0.028$ & $0.201858\pm0.001502$ & $0.201860\pm0.001215$ \\
Power law ($\theta_{\rm PL}$): MPS2 prior &
$1.144\pm0.027$ & $0.187428\pm0.001432$ & $0.187430\pm0.001159$ \\
Power law ($\theta_{\rm PL}$): SATLAS prior &
$1.156\pm0.028$ & $0.279457\pm0.001540$ & $0.279459\pm0.001245$ \\
\shortstack[l]{Fixed CLV ($\theta_{\rm CLV}$): PHOENIX\\[-1pt]node} &
$1.151\pm0.025$ & --- & --- \\
\shortstack[l]{SB-consistent PHOENIX\\[-1pt]($\theta_{\rm LD,PHX}$)} &
$1.154\pm0.026$ & --- & --- \\
Fixed CLV ($\theta_{\rm CLV}$): Stagger &
$1.160\pm0.028$ & --- & --- \\
Fixed CLV ($\theta_{\rm CLV}$): Kurucz &
$1.140\pm0.027$ & --- & --- \\
Fixed CLV ($\theta_{\rm CLV}$): MPS1 &
$1.149\pm0.028$ & --- & --- \\
Fixed CLV ($\theta_{\rm CLV}$): MPS2 &
$1.147\pm0.028$ & --- & --- \\
Fixed CLV ($\theta_{\rm CLV}$): SATLAS &
$1.165\pm0.028$ & --- & --- \\
\bottomrule
\end{tabular}
\begin{minipage}{0.95\textwidth}
\footnotesize\textit{Notes.}
The table contains only the joint five-epoch fit. In the second column,
$\theta_{\rm fit}$ is the fitted angular diameter in milliarcseconds. In the
third column, $\alpha_{\rm prior}$ is the Gaussian-prior mean and width,
written as $\mu\pm\sigma$. In the fourth column, $\alpha_{\rm post}$ is the
posterior 68\% interval for the fitted limb-darkening exponent $\alpha$.
The fitted diameter is $\theta_{\rm UD}$ for the uniform-disk row, $\theta_{\rm PL}$ for the
power-law rows, and $\theta_{\rm CLV}$ for the fixed-CLV rows. 
Their diameter uncertainties include the fit bootstrap and wavelength-scale terms. 
MPS1 and MPS2 denote the two MPS-ATLAS atmosphere grids. Fit definitions and
atmosphere-grid coverage are given in Section~\ref{sec:data_analysis}.
\end{minipage}
\end{table*}

\begin{table*}[t]
\centering
\caption{Adopted inputs, interferometric measurements, and derived properties of Proxima Centauri compared with Rabus et al.~(2019).}
\label{tab:target-properties}
\scriptsize
\setlength{\tabcolsep}{5pt}
\begin{tabular}{llcc}
\toprule
Quantity & Role in this analysis & Value & \citet{Rabus2019} \\
\midrule
$T_{\rm eff, input}$ & Atmosphere input & $2992^{+49}_{-47}$~K & --- \\
$\log g_{\rm input}$ & Atmosphere input & $5.20\pm0.23$~dex & --- \\
$[\mathrm{Fe/H}]_{\rm input}$ & Atmosphere input & $+0.21$~dex & --- \\
Distance & Gaia EDR3 (corrected) & $1.30193\pm0.00008$~pc & $1.30124\pm0.00034$~pc \\
$F_{\rm bol}$ & Adopted; SED check & $(2.86\pm0.14)\times10^{-8}$~erg~s$^{-1}$~cm$^{-2}$ & $(2.866\pm0.210)\times10^{-8}$~erg~s$^{-1}$~cm$^{-2}$ \\
$\theta_{\rm UD}$ & Interferometric fit & $1.117\pm0.024$~mas & $1.066\pm0.007$~mas \\
$\theta_{\rm LD,PHX}$ & PHOENIX fit & $1.154\pm0.026$~mas & $1.103\pm0.007$~mas \\
$R_\star$ & Derived & $0.162\pm0.004\,R_\odot$ & $0.154\pm0.001\,R_\odot$ \\
$T_{\rm eff,SB}$ & Derived & $2833\pm49$~K & $2901\pm68$~K \\
\bottomrule
\end{tabular}
\begin{minipage}{0.95\textwidth}
\footnotesize\textit{Notes.} The atmosphere parameters in the first three rows are inputs to the PHOENIX calculation. 
The distance is adopted from Gaia EDR3 after applying the Lindegren et al.~(2021) zero-point correction. 
The bolometric flux is adopted from Ribas et al.~(2017).
The Rabus et al.~distance is obtained from their quoted parallax of $768.500\pm0.203$~mas. 
Their $T_{\rm eff}=2901\pm68$~K is their derived temperature, whereas $T_{\rm eff,SB}=2833$~K is derived here from the adopted flux and fitted diameter. 
The quoted direct-fit diameter uncertainties combine bootstrap and wavelength-scale terms. 
A single total uncertainty is reported for each PHOENIX-derived quantity. 
The $0.160-0.163~R_\odot$ range from the alternative atmosphere grids is reported separately and is not included in the adopted uncertainty.
\end{minipage}
\end{table*}

\section{Observations, Data Reduction, and Analysis}\label{sec:observations}
We used five $H$-band PIONIER epochs: two from 2014 that were previously published by \citet{Rabus2019,Lachaume2019}, and three unpublished epochs from 2017 and 2019. The ESO programme IDs, observing configurations, and calibrators are listed in Table~\ref{tab:calibrators}. PIONIER is a four-telescope interferometric combiner \citep{LeBouquin2011} at the VLTI, Paranal, Chile. For all five epochs, we downloaded the raw science and calibrator data from the ESO archive and reduced them with \texttt{pndrs} (version 3.94). Details of the transfer-function calibrations  are given in Appendix~\ref{sec:transfer-function}. Except for HD~128713 in 2017, the calibrators are effectively unresolved over our sampled baselines, with calibrated $V^2$ values close to unity. The partial resolution of HD~128713 is accounted for through its predicted visibility in the transfer-function calibration. Its larger angular diameter and associated uncertainty increase the propagated calibration uncertainty, but we find no measurable change in Proxima's fitted diameter. The smaller calibrator HD~125640 was also observed on the same night, so the calibration does not rely on HD~128713 alone.

\begin{figure*}
\centering
\includegraphics[width=0.45\textwidth]{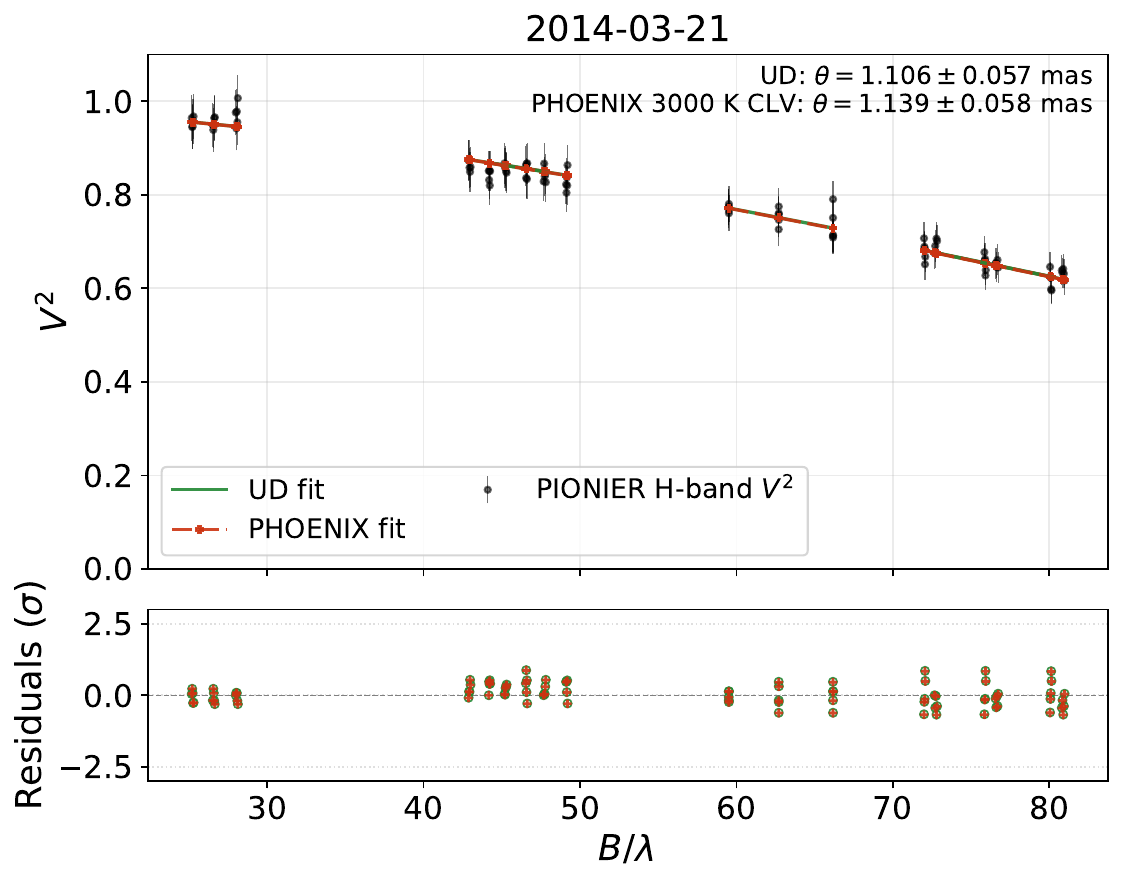}
\includegraphics[width=0.45\textwidth]{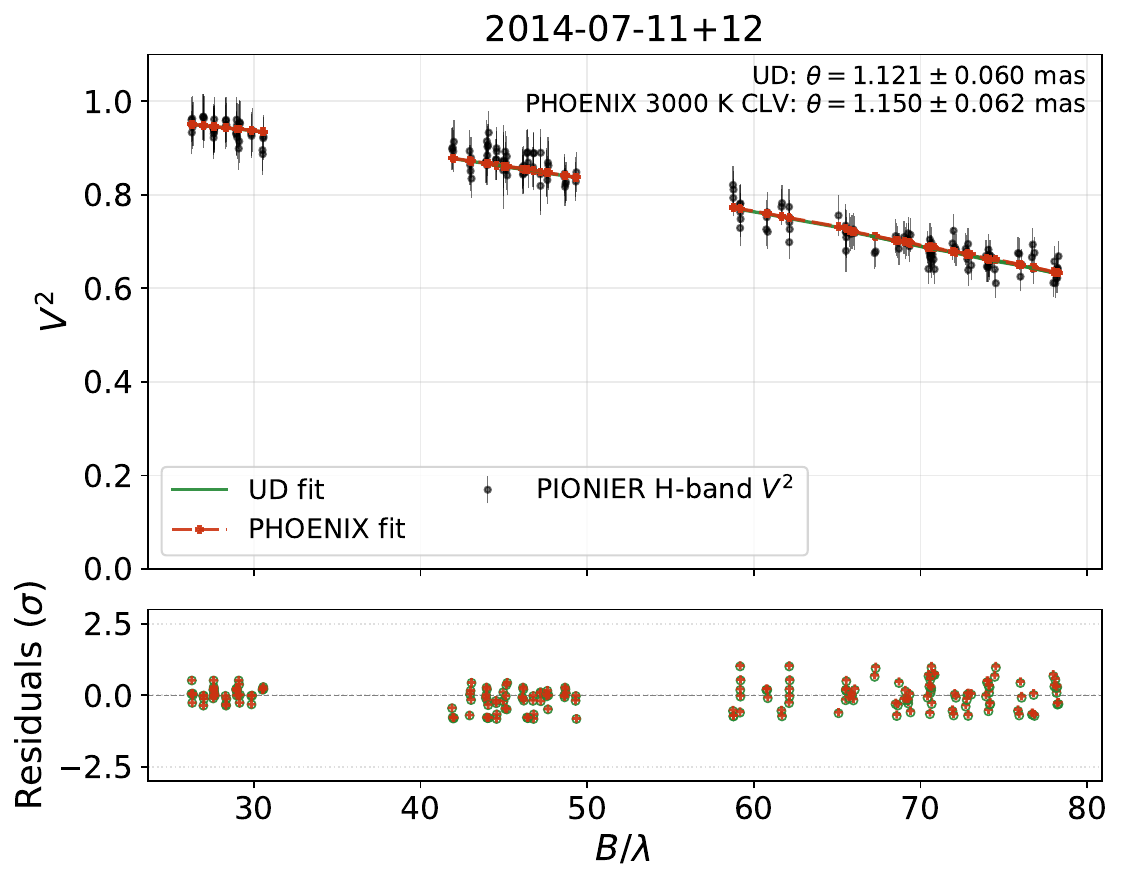}\\
\includegraphics[width=0.45\textwidth]{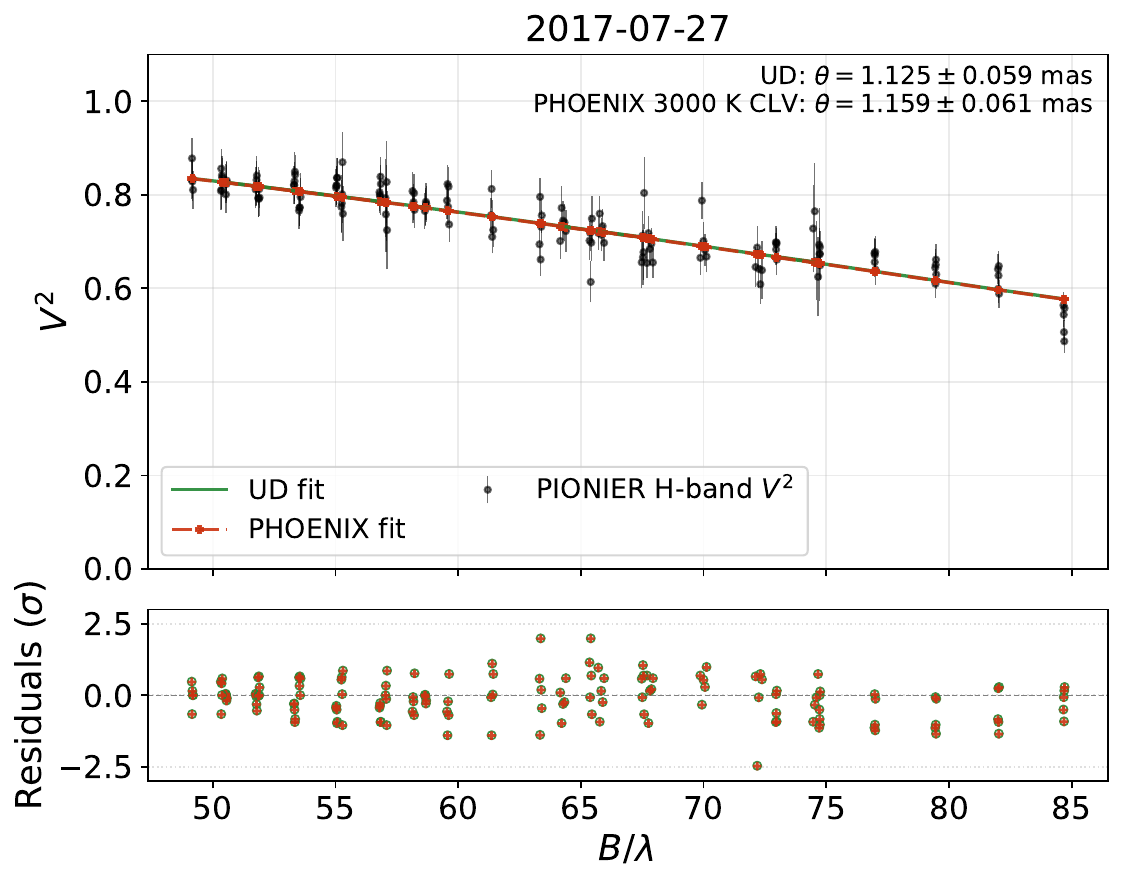}
\includegraphics[width=0.45\textwidth]{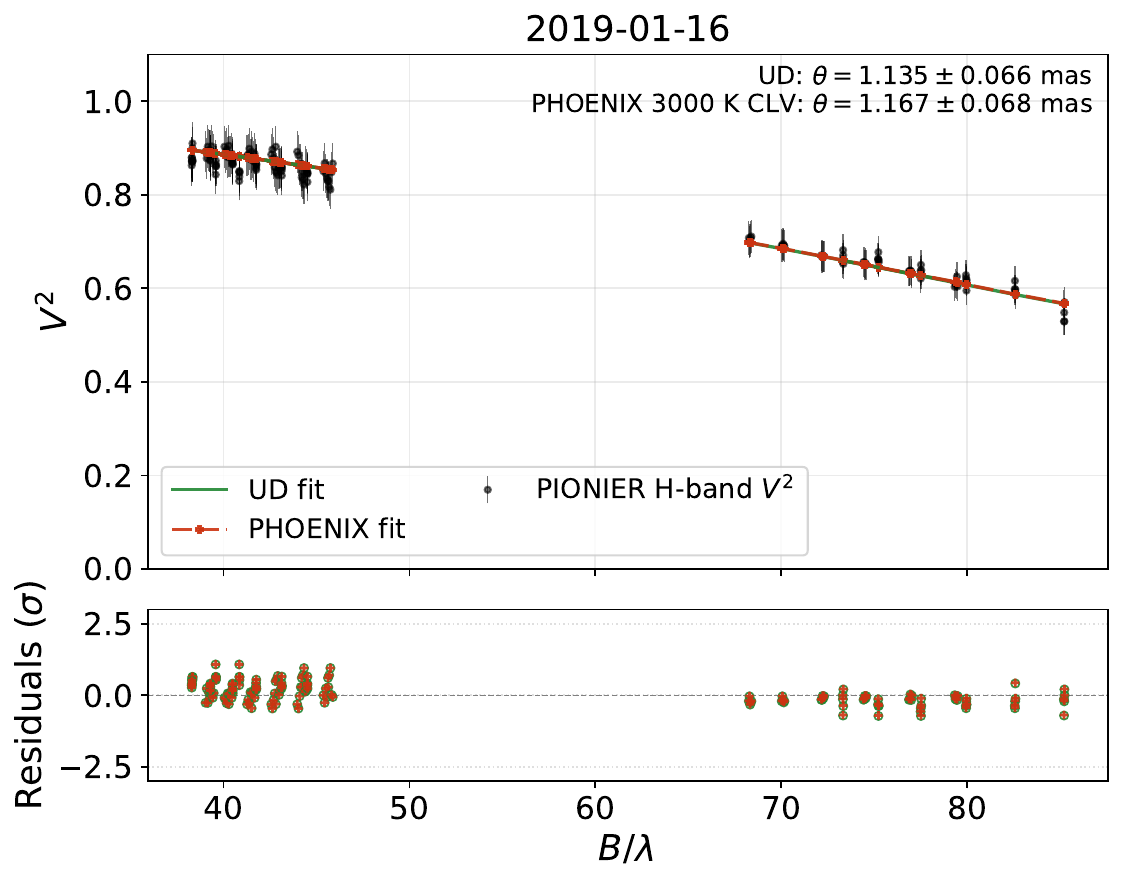}\\
\includegraphics[width=0.45\textwidth]{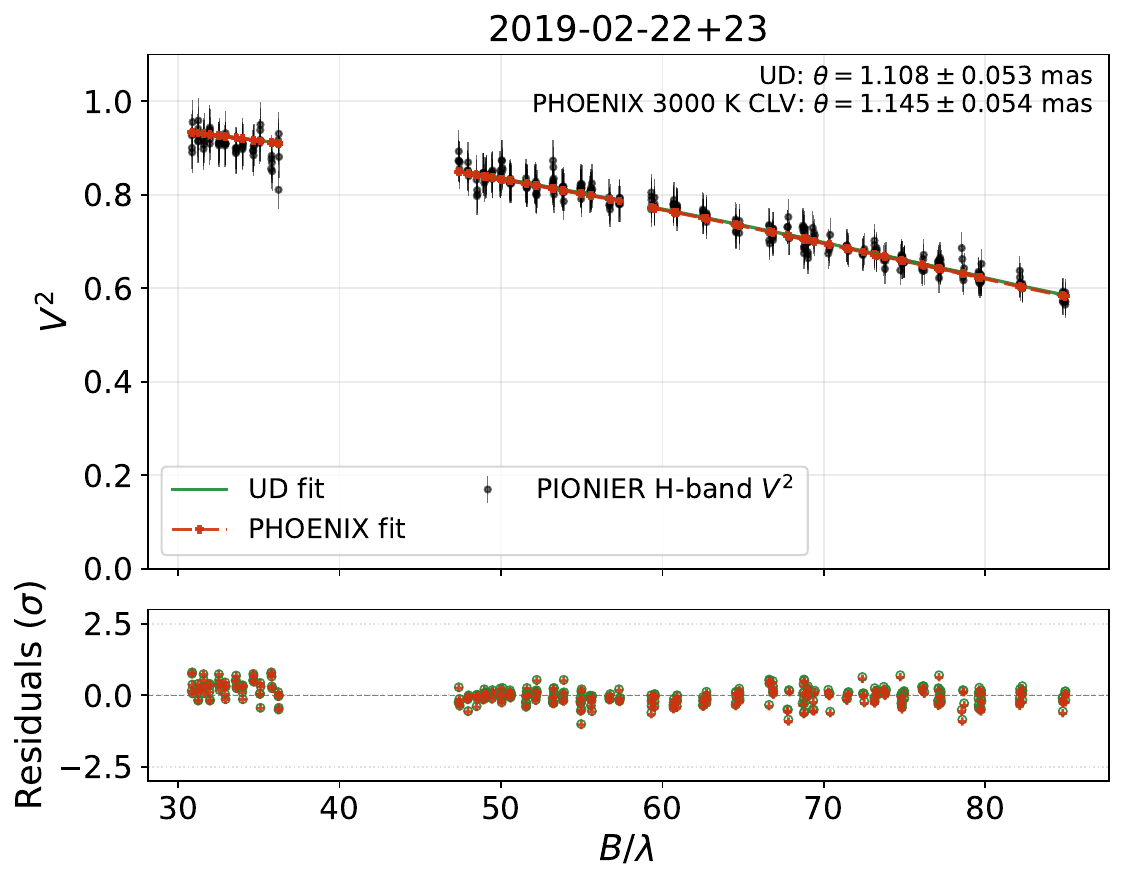}
\caption{Individual calibrated PIONIER H-band squared visibilities and residuals for the five epochs listed in Table~\ref{tab:calibrators}. The curves show the uniform-disk and representative PHOENIX CLV fits to each data set. Each annotation gives the fit and its bootstrap uncertainty. From left to right and top to bottom, the panels show 2014 March, 2014 July, 2017 July, 2019 January, and 2019 February 22--23.}
\label{fig:v2}
\end{figure*}

\begin{figure}
\centering
\includegraphics[width=0.47\textwidth]{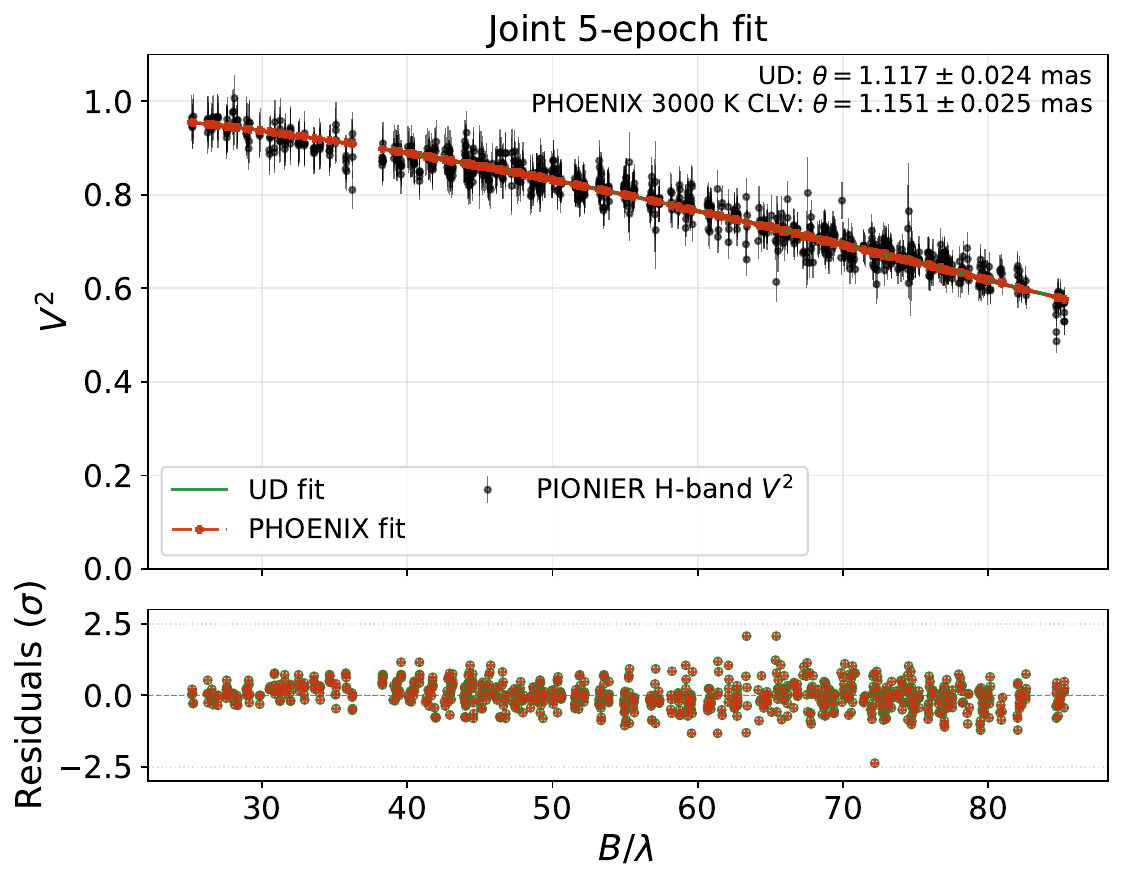}
\caption{Joint five-epoch calibrated PIONIER $H$-band squared visibilities and residuals. All five observations are fitted with one common angular diameter. The UD and PHOENIX annotations report the fits and the uncertainties listed in Table~\ref{tab:diameter-fits}; the displayed PHOENIX value, $1.151\pm0.025$~mas, is the fixed 3000~K PHOENIX-node fit.
The adopted PHOENIX diameter, obtained by interpolating the 2800 and 2900 K fits until the Stefan–Boltzmann temperature matches the model temperature, is $1.154\pm0.026$ mas (Table~\ref{tab:diameter-fits}). 
The corresponding individual epoch fits are shown in Figure~\ref{fig:v2}, while the full joint atmosphere-grid variations are listed in Table~\ref{tab:diameter-fits}.}
\label{fig:v2_combined}
\end{figure}

All observations were obtained in the $H$ band, $1.51$--$1.76~\mu$m, using Auxiliary Telescopes (ATs). The 2014 data comprise three spectral channels, while the 2017 and 2019 data comprise six channels. Together with the different baseline configurations, these observations provide complementary spatial-frequency coverage. The calibrated squared visibilities and diameter fits are shown in Figures~\ref{fig:v2} and \ref{fig:v2_combined}.

We checked the calibrator closure phases, which are sums of the fringe phases measured around a three-telescope triangle; they show no significant departure from centro-symmetry. We also searched the Gaia EDR3 proper-motion-anomaly catalog of \citet{Kervella2022}. These checks provide no evidence for a companion.

For completeness, we considered additional archival observations of Proxima but excluded them from the present analysis. The 2016 PIONIER observations relied on a calibrator with an angular diameter of approximately 1.8\,mas, making them poorly suited to calibrating this partially resolved target. The 2017 August 27 PIONIER observations contain only one calibrator and exhibit unexpectedly low $V^2$ values at small spatial frequencies, while the 2019 January 23 observations cannot be calibrated because of technical issues. We also reduced 2017 GRAVITY data obtained under ESO programs 299.C-5039 and 099.C-0865 (PI Kervella), but did not include them in the diameter fits because their calibrated $V^2$ uncertainties were approximately $10\%$. The retained 2019 PIONIER programme is also represented in the survey of \citet{Rains2020}; however, Proxima was not included in that sample, so the survey provides observing-programme context rather than an independent Proxima measurement.

The OIFITS files and data-analysis scripts are deposited in Zenodo~\citep{anugu_2026_proxima_zenodo};  OIFITS files are also available through the optical interferometry database OIDB\footnote{\url{https://oidb.jmmc.fr}}.

\section{Data fitting and analysis methods}\label{sec:data_analysis}

\subsection{Squared Visibility Modeling and Uncertainty Estimation}

We modeled the calibrated $V^2$ from all five epochs with \texttt{PMOIRED}\footnote{\url{https://github.com/amerand/PMOIRED}} \citep{Merand2022PMOIRED}. We fit a uniform-disk diameter as the reference angular scale, then use power-law and atmosphere-based center-to-limb-variation (CLV) models to estimate the limb-darkening correction (Figures~\ref{fig:v2} and \ref{fig:v2_combined}).

We fitted a power-law limb-darkened disk with intensity profile $I(\mu)/I(1)=\mu^\alpha$, where $\mu$ is the cosine of the emergent angle. Following \citet{Anugu2026LimbDarkening}, we fitted the angular diameter $\theta_{\rm PL}$ and limb-darkening exponent $\alpha$ jointly to the PIONIER $H$-band data. For Proxima's adopted parameters—$T_{\rm eff}=2992^{+49}_{-47}\,\mathrm{K}$ \citep{Pineda2021}, $\log g=5.20\pm0.23$ \citep{Segransan2003}, and $[\mathrm{Fe/H}]=+0.21$ \citep{Schlaufman2010}—we used  a python based \texttt{ExoTiC-LD}\footnote{\url{https://github.com/nanugu/ExoTiC-LD.git}, \url{https://exotic-ld.readthedocs.io/en/latest/views/supported_stellar_grids.html}} \citep{Grant2024JOSS} software to calculate bandpass-integrated predictions for $\alpha$ and its uncertainty from the PHOENIX \citep{Husser2013}, Kurucz \citep{Kurucz1993}, Stagger \citep{Magic2013,Magic2015}, and MPS-ATLAS \citep{Kostogryz2022} grids. We obtained a corresponding SATLAS prediction by fitting its tabulated spherical intensity profile \citep{Neilson2013}\footnote{\url{http://cdsarc.u-strasbg.fr/viz-bin/qcat?J/A+A/554/A98}}. Each predicted coefficient and uncertainty define the mean and width of a Gaussian prior on $\alpha$; $\alpha$ remains free to vary in the visibility fit. Table~\ref{tab:diameter-fits} lists the prior and fitted $\alpha$ values. 

We use PHOENIX as  our reference because its grid covers Proxima’s adopted $T_{\rm eff}$, $\log g$, and $[\mathrm{Fe/H}]$. Our calculation used ExoTiC-LD software git commit \texttt{1bcce41} and selected the nearest grid nodes without extrapolation.

Finally, we fitted the $V^2$ data using fixed, bandpass-integrated CLV profiles from the atmosphere grids (see Figure~\ref{fig:v2}). A CLV profile describes how the stellar intensity changes from the center of the disk to the limb. We use this tabulated intensity profile directly, avoiding an analytic limb-darkening parameterization, and fit only the angular diameter, $\theta_{\rm CLV}$. Fixing the profile removes the need to fit independent limb-darkening coefficients and their degeneracy with diameter, although the inferred diameter remains conditional on the adopted atmosphere profile. Table~\ref{tab:diameter-fits} summarizes the joint UD, prior-constrained power-law, and fixed-CLV fits.

Because Proxima is only partially resolved, formally small errors on individual $V^2$ points would allow a few points to dominate the fit. We therefore impose a conservative minimum relative uncertainty of 5\% in $V^2$ to represent residual calibration correlations; this floor is applied in every fit and bootstrap. We used 500 refits. For each fit, we formed a new data set by randomly selecting the number of epochs, observing-time blocks in modified Julian date (MJD), and physical baselines (about 37\% are omitted on average). All wavelength channels from each selected observation were kept together. Following \citet{Gallenne2018,Gallenne2019} and \citet{Lachaume2019}, we varied the wavelength scale independently for each epoch by 0.35\%, using one scale for all channels in that epoch. The quoted uncertainty is the standard deviation of 500 refits that resample the observations and vary the wavelength scale.

\subsection{Stefan--Boltzmann $T_{\rm eff, SB}$ Calculation}\label{sec:sb-temperature}

We derive the effective temperature $T_{\rm eff,SB}$ from the joint five-epoch fit using the Stefan--Boltzmann relation. We adopt the broad spectral-energy distribution integrated by \citet{Ribas2017}, which gives
$F_{\rm bol}=(2.86\pm0.14)\times10^{-8}$~erg~s$^{-1}$~cm$^{-2}$, as the primary flux for the Stefan--Boltzmann calculation. For comparison, \citet{Rabus2019} reported
$F_{\rm bol}=(2.866\pm0.210)\times10^{-8}$~erg~s$^{-1}$~cm$^{-2}$. The central values agree, although the \citet{Rabus2019} uncertainty is larger and their PHOENIX SED analysis used priors on the temperature, radius, distance, and angular diameter. Several studies have documented mismatches between synthetic and observed M-dwarf spectra and flux distributions, particularly at optical wavelengths \citep{Reyle2011,Mann2013,McKemmish2019,Jones2023,Rains2024}. As an independent check, we fitted the available broadband photometry with \textsc{astroARIADNE} \citep{Vines2022ARIADNE}, using several atmosphere grids and Bayesian model averaging. This fit gives $F_{\rm bol}=2.884^{+0.283}_{-0.251}\times10^{-8}$~erg~s$^{-1}$~cm$^{-2}$, consistent with the Ribas et al. value. We retain the Ribas et al. flux for the primary Stefan--Boltzmann calculation.

For each PHOENIX grid temperature $T_j$, we generated an $H$-band CLV profile, fitted only its angular diameter $\theta_{{\rm CLV},j}$ to the PIONIER $V^2$ data, and used that diameter and $F_{\rm bol}$ to calculate
\begin{equation}
 T_{{\rm SB},j}=\left(\frac{4F_{\rm bol}}{\sigma_{\rm SB}\theta_{{\rm CLV},j}^2}\right)^{1/4},
\end{equation}
where $\theta_{{\rm CLV},j}$ is in radians and $\sigma_{\rm SB}$ is the Stefan--Boltzmann constant. For each PHOENIX grid temperature $T_j$, we calculate $T_{{\rm SB},j}$ from the fitted diameter. We then interpolate between grid points to find the temperature at which $T_{{\rm SB},j}=T_j$; we denote this final self-consistent value by $T_{\rm eff,SB}$. We call the corresponding interpolated diameter $\theta_{\rm LD,PHX}$. We recorded the PIONIER reduced $\chi^2$ at each grid point but did not add it to the temperature comparison. The fixed SATLAS profile is spherical, whereas the \texttt{ExoTiC-LD} profiles are plane-parallel. The adopted stellar parameters, distance, bolometric flux, and primary derived properties are summarized in Table~\ref{tab:target-properties}.

\section{Results}\label{sec:results}

\subsection{Joint Uniform-Disk Fit}

Fitting the five data sets simultaneously with one common angular scale gives
$\theta_{\rm UD,joint}=1.117\pm0.024\ {\rm mas}$. The calibrated visibilities and residuals are shown in Figures~\ref{fig:v2} and \ref{fig:v2_combined}.

For comparison, \citet{Rabus2019} and \citet{Lachaume2019} reported $\theta_{\rm UD}=1.066\pm0.007$~mas from the same 2014 campaign; \citet{Lachaume2019} used a correlated-error fit. Our five-epoch value is $0.051$~mas larger, corresponding to a $2.0\sigma$ difference when the quoted uncertainties are combined in quadrature.

\subsection{Fixed-CLV Fitted Diameters and $T_{\rm eff,SB}$}\label{sec:fixed-clv-results}

We compared the data with PHOENIX profiles that include limb darkening, using the adopted stellar parameters (Table~\ref{tab:target-properties}). The input temperature is $2992$~K, so the nearest PHOENIX grid profile is $3000$~K. Keeping this profile fixed and fitting only the diameter gives $\theta_{\rm CLV,PHX}=1.151\pm0.025$~mas. We then repeated the fit with the 2800 and 2900~K profiles and used each fitted diameter and $F_{\rm bol}$ to calculate a temperature. The 2800~K profile gives $\theta_{\rm CLV,PHX}=1.155\pm0.026$~mas and $T_{\rm SB}=2832\pm47$~K, while the 2900~K profile gives $\theta_{\rm CLV,PHX}=1.153\pm0.026$~mas and $T_{\rm SB}=2835\pm47$~K. These fits bracket the point where the calculated temperature matches the PHOENIX profile temperature. Interpolating between them gives
$T_{\rm eff,SB} = 2833\pm49\ {\rm K}$,
$\theta_{\rm LD,PHX} = 1.154\pm0.026\ {\rm mas}$.

We correct the Gaia EDR3 parallax using the source-dependent zero-point offset
$\Delta\varpi_{\rm zp}=-0.022304$~mas following \citet{Lindegren2021}. We use
$\varpi_{\rm corr}=\varpi_{\rm cat}-\Delta\varpi_{\rm zp}$, where
$\varpi_{\rm cat}$ is the catalog parallax. This gives
$\varpi=768.089\pm0.050$~mas and $d=1.30193$~pc \citep{Gaia2021}. Here,
$\theta_{\rm LD}$ is the adopted limb-darkened angular diameter in mas and
$d$ is the distance in parsecs. We calculate
\begin{equation}
 \frac{R_\star}{R_\odot}=\frac{\theta_{\rm LD}({\rm mas})\,d({\rm pc})}{9.30094}.
\end{equation}
Using the joint Stefan--Boltzmann-consistent PHOENIX diameter and the Gaia distance, we infer $R_\star=0.162\pm0.004\,R_\odot$. 
The alternative fixed-CLV atmosphere models give radii from $0.160$ to $0.163\,R_\odot$. We report this range to show the model-to-model variation; it is not added to the uncertainty of the adopted PHOENIX radius.
This radius is larger than $0.154\,R_\odot$ inferred from the PIONIER result of \citet{Rabus2019}.

We report the standard deviation of 500 refits for each fixed-CLV PHOENIX diameter. These refits resample the observations and vary the wavelength scale. The Stefan--Boltzmann-consistent PHOENIX value uses the $0.026$~mas uncertainty adopted in the interpolation. The limb-darkened diameter therefore depends on the adopted atmosphere model; it is not measured directly from the interferometric data.

The uncertainty in $T_{\rm eff,SB}$ includes contributions from $F_{\rm bol}$ and $\theta_{\rm LD,PHX}$. At either grid point, these contribute approximately 34 and 32~K, respectively; the displayed values are rounded, and the unrounded contributions combine in quadrature to the local uncertainty. Dividing that unrounded value by the measured interpolation slope, $|d(T_{\rm SB}-T_{\rm grid})/dT_{\rm grid}|=0.972$, gives 48.5~K, rounded to the reported $\pm49$~K.


\begin{figure*}
\centering
\includegraphics[width=\textwidth]{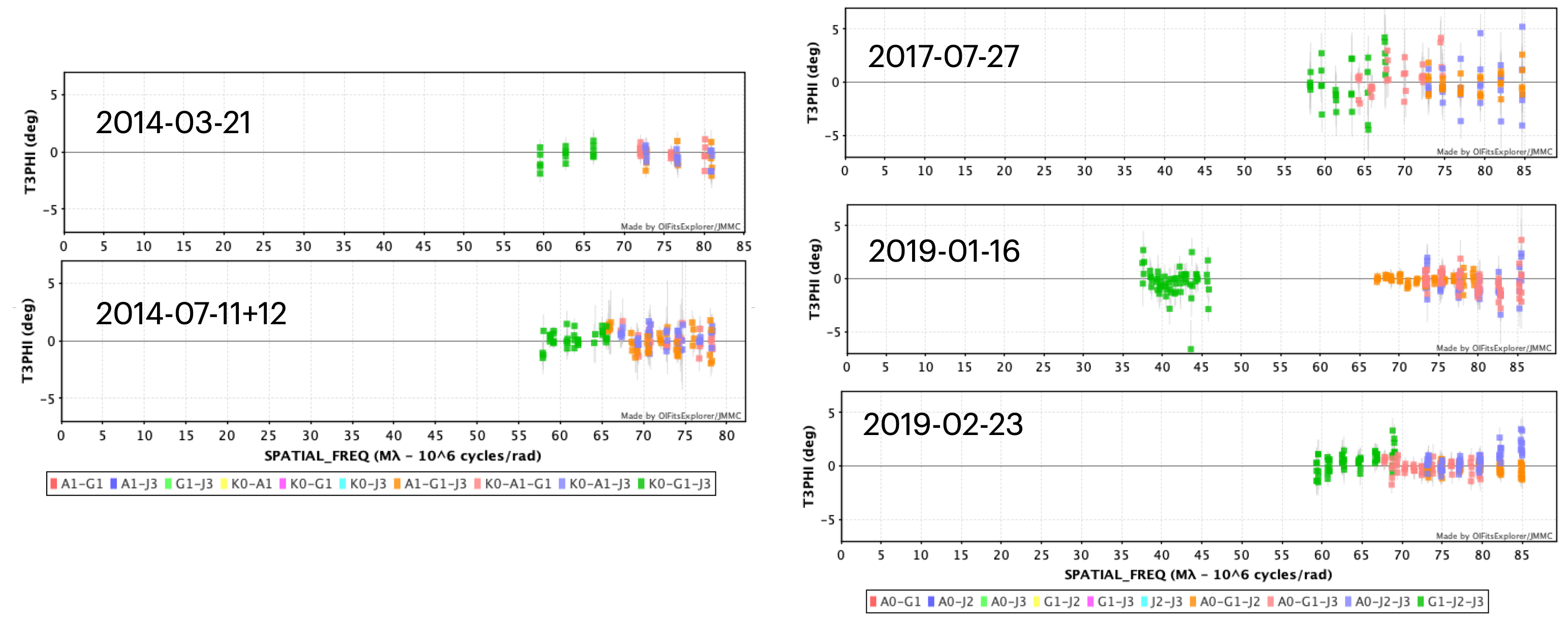}
\caption{Calibrated closure phases for the five epochs listed in Table~\ref{tab:calibrators}. The closure phase (T3PHI) is plotted as a function of spatial frequency; colors identify the corresponding telescope triangle. The horizontal line marks zero closure phase within the uncertainties.}
\label{fig:closure-phases}
\end{figure*}

\begin{figure*}
\centering
\includegraphics[width=0.60\textwidth]{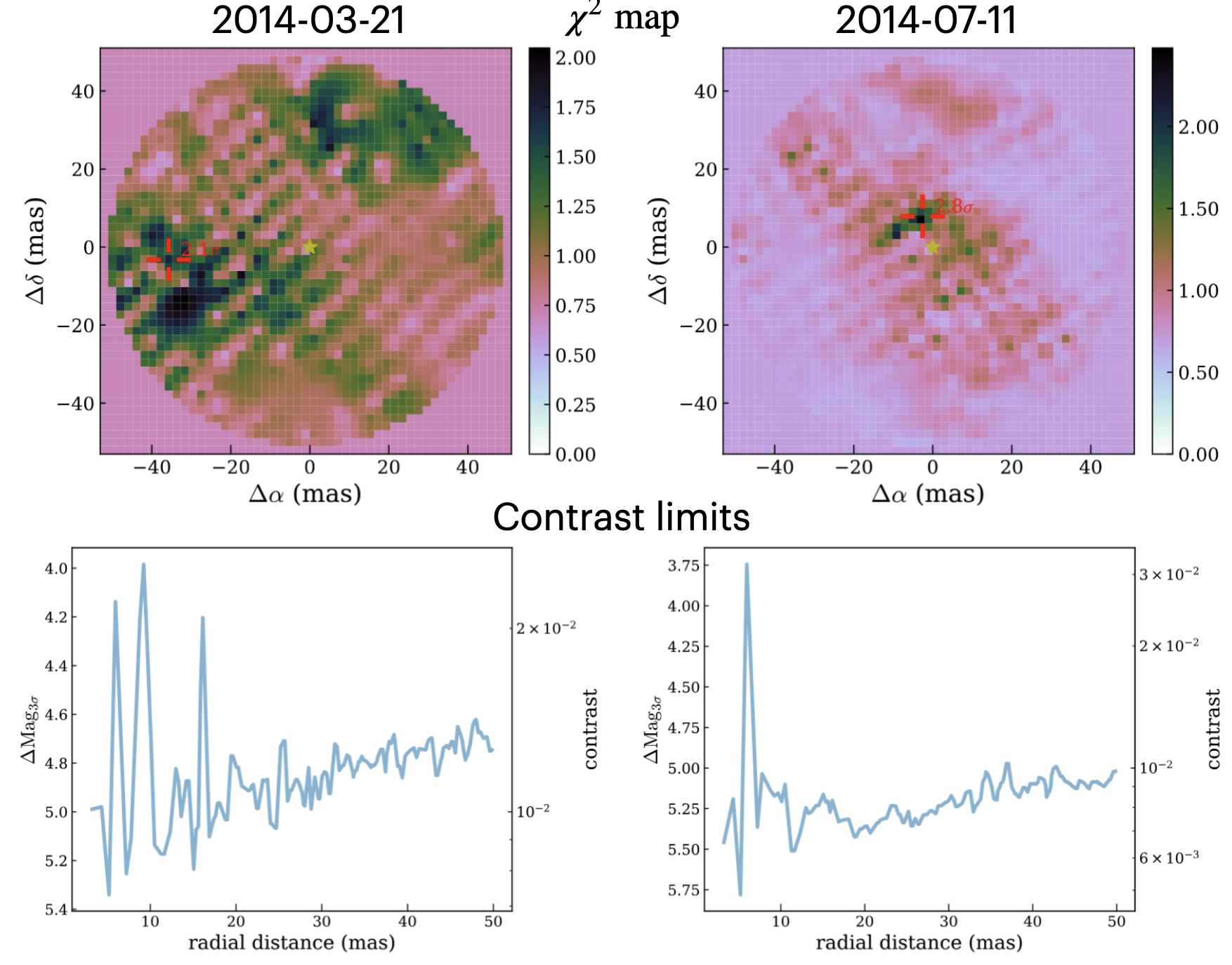}
\includegraphics[width=0.92\textwidth]{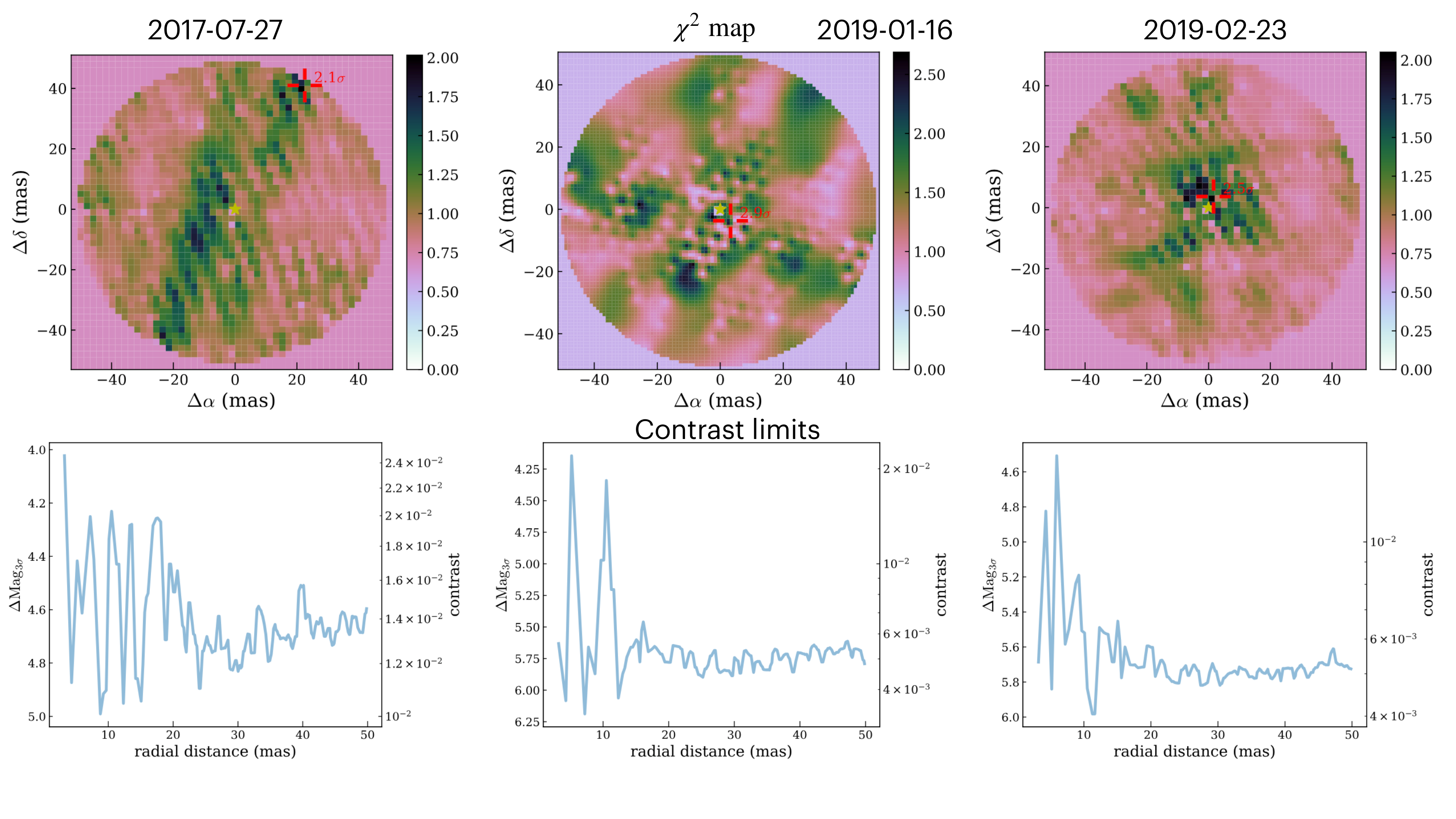}
\caption{\texttt{CANDID} companion-search results for the five epochs listed in Table~\ref{tab:calibrators}, using joint $V^2$ and closure-phase fits. The upper panels show reduced-$\chi^2$ maps (the color scale is reduced $\chi^2$), and the lower panels show radial $3\sigma$ contrast limits from 500 random-position injection-recovery trials. The \texttt{CANDID} and \texttt{PMOIRED} searches gave the same result, and no peak reaches $3\sigma$.}
\label{fig:companion-limits}
\end{figure*}

\subsection{Companion Search and Closure-Phase Sensitivity}

Previous searches for Proxima companions have used direct imaging, radial velocities, and astrometric monitoring, probing orbital and separation regimes different from those accessible to long-baseline interferometry \citep{Benedict1999,Lurie2014,Beamin2017}. For example, the HST astrometry and radial velocities of \citet{Benedict1999} excluded companions above $0.8\,M_{\rm Jup}$ for periods of 1--1000 days, while the VVV imaging survey constrained much wider companions around Proxima \citep{Beamin2017}. An early HST feature interpreted as a possible substellar companion was not confirmed by subsequent WFPC2 images \citep{Golimowski1998}. The published PIONIER diameter study of \citet{Rabus2019} used Proxima but did not present a target-specific binary-grid search or closure-phase contrast map.

We performed the companion search using closure phases (Figure~\ref{fig:closure-phases}) with \texttt{CANDID} \citep{Gallenne2015} and \texttt{PMOIRED} using binary-grid fits and injection-recovery tests (Figure~\ref{fig:companion-limits}). We fit the primary uniform-disk diameter, companion position $(\Delta\alpha,\Delta\delta)$, and companion-to-primary $H$-band flux ratio $f$, fixing the primary flux to unity and treating the companion as unresolved. We restricted $f<0.1$. For the \texttt{PMOIRED} search, we defined the angular scale $s=[\max(B_{\rm p}/\lambda)]^{-1}$, where $B_{\rm p}$ is the projected baseline, and the mean spectral resolving power $\mathcal{R}=\langle\lambda/\Delta\lambda\rangle$, where $\Delta\lambda$ is the spectral-channel width. We used $s$ as the initial grid spacing and adopted the inner and outer search radii $\rho_{\min}=s/2$ and $\rho_{\max}=\mathcal{R}s/2$. The inner radius is an adopted angular-resolution cutoff, while the outer radius limits the search to the effective interferometric field of view set by spectral coherence; bandwidth smearing increasingly attenuates a companion's signal at larger separations \citep{Gallenne2015}. For the 2019 February epoch, $s=2.43$~mas and $\mathcal{R}=34.27$, giving $1.21<\rho\leq41.60$~mas over a $[-41.60,41.60]$~mas grid. These limits depend on the epoch's spatial-frequency coverage and spectral resolution. The \texttt{CANDID} maps in Figure~\ref{fig:companion-limits} extend to approximately 50~mas. Both searches yield consistent non-detections within their sampled regions.

Radial contrast limits are based on 500 random-position injection-recovery trials. We adopted the conventional $3\sigma$ local detection threshold used in interferometric companion searches \citep{Gallenne2015}.  The highest local peaks are approximately $2.1\sigma$, $2.8\sigma$, $2.1\sigma$, $2.9\sigma$, and $2.5\sigma$ in the 2014 March, 2014 July, 2017 July, 2019 January, and 2019 February epochs, respectively. They occur at different positions, and no peak reaches this threshold (see Figure~\ref{fig:companion-limits}). For the three later epochs, the approximate contrast limits are $\Delta H\simeq4.6$~mag ($f\simeq0.014$), $\Delta H\simeq5.7$~mag ($f\simeq0.005$), and $\Delta H\simeq5.6$~mag ($f\simeq0.006$), respectively. This work constrains only the sampled $H$-band angular field of view and does not exclude companions at other separations, flux ratios, wavelengths, or orbital phases.

\section{Discussion}\label{sec:discussion}

We first assess whether the angular-diameter offset is robust across epochs, reductions, and calibrator choices, and then examine whether the implied stellar radius and effective temperature are physically plausible.

\subsection{Stellar radius}
The joint fit to all five epochs yields $\theta_{\rm UD}=1.117\pm0.024$~mas, 4.8\% larger than the $\theta_{\rm UD}=1.066\pm0.007$~mas reported by \citet{Rabus2019}. Combining the quoted uncertainties in quadrature gives a difference of $2.0\sigma$. Similarly, our PHOENIX diameter, $\theta_{\rm LD,PHX}=1.154\pm0.026$~mas, is 4.6\% larger than their $\theta_{\rm LD}=1.103\pm0.007$~mas, a difference of $1.9\sigma$. The same difference appears in separate fits to the \citet{Rabus2019} used two 2014 epochs, which yield $\theta_{\rm UD}=1.106\pm0.057$~mas for 2014 March 21 and $1.121\pm0.060$~mas for 2014 July 11. The three later epochs yield diameters in the same range, indicating that the larger joint diameter is not driven by a single epoch (Figure~\ref{fig:v2}).

The adopted angular diameters of the principal calibrators agree with those used by \citet{Rabus2019} and \citet{Lachaume2019}. Given this agreement, the remaining $1.9\sigma$ offset may reflect differences in data reduction, fitting, or uncertainty treatment.

For historical comparison, applying the corrected Gaia distance to the
reported limb-darkened VINCI (2002 Feb) and AMBER (2008 Feb) diameters gives radii of
approximately $0.143\pm0.011\,R_\odot$ and
$0.142\pm0.007\,R_\odot$, respectively
\citep{Segransan2003,Demory2009}. Our five-epoch PIONIER analysis gives
$R_\star=0.162\pm0.004\,R_\odot$, nominally $1.6\sigma$ and
$2.5\sigma$ larger than the VINCI and AMBER values, respectively.

Our inferred radius is also consistent with those of the nearly equal-mass, rapidly rotating, fully convective dwarfs GJ~65 A and B, $R_A=0.165\pm0.006\,R_\odot$ and $R_B=0.159\pm0.006\,R_\odot$. Their respective masses are $0.1225\pm0.0043\,M_\odot$ and $0.1195\pm0.0043\,M_\odot$ \citep{Kervella2016}. \citet{Kervella2016} interpreted their inflated radii as possible evidence for magnetic inhibition of convection associated with rapid rotation and strong magnetic fields. Proxima, with $M_\star=0.1221\pm0.0022\,M_\odot$ \citep{Kervella2017}, also exhibits an approximately seven-year activity cycle \citep{Wargelin2017}, raising the possibility that magnetic activity could affect either its physical radius or its apparent interferometric diameter. However, our 2014, 2017, and 2019 measurements show no significant epoch-to-epoch diameter variation. Long-term interferometric observations accompanied by contemporaneous activity monitoring would be required to test any connection.

\subsection{Physical plausibility of the inferred parameters}

Assuming coevality with $\alpha$ Cen A and B, we adopt an age of $4.8~\mathrm{Gyr}$ for Proxima \citep{Bazot2016,Ribas2016}.
At Proxima’s estimated mass, $M_\star=0.1221\pm0.0022\,M_\odot$ \citep{Kervella2017}, interpolation of the solar-metallicity BHAC15 tracks \citep{Baraffe2015} yields $R_\star\simeq0.148\,R_\odot$. Our inferred radius, $R_\star=0.162\pm0.004\,R_\odot$, is larger by approximately $0.014\,R_\odot$, or 10\%. This comparison remains qualitative because the BHAC15 tracks do not sample Proxima's super-solar metallicity and we do not assign a systematic uncertainty to the model prediction \citep{Ribas2016}.

The derived $T_{\rm eff,SB}=2833\pm49$~K is about $147$~K below the
$2980\pm80$~K estimate of \citet{Ribas2017}; combining the quoted errors alone
gives a descriptive difference of about $1.6\sigma$. On the Pecaut--Mamajek temperature scale, our lower temperature would place
Proxima near M6V rather than the commonly adopted M5.5V classification
\citep[][see Mamajek's updated online compilation\footnote{\url{https://github.com/emamajek/SpectralType}}]{PecautMamajek2013}.

Using the 2MASS $K_s$ magnitude \citep{Skrutskie2006} and the corrected Gaia distance,
the empirical radius--$M_{K_s}$ relation of \citet{Mann2015} predicts
$R_\star=0.154\pm0.005\,R_\odot$. The corresponding angular diameter is
$\theta_{\rm LD}=1.10\pm0.04$~mas. Using the quoted uncertainties, these
values are consistent with our
$R_\star=0.162\pm0.004\,R_\odot$ and
$\theta_{\rm LD,PHX}=1.154\pm0.026$~mas at
$1.2\sigma$ and $1.1\sigma$, respectively. The comparison is not fully
independent because both radius estimates use the corrected Gaia
distance, while the Mann et al. calibration was based on radii inferred
from bolometric fluxes, parallaxes, and empirically calibrated
spectroscopic effective temperatures. Using the adopted $F_{\rm bol}$ and
the corresponding angular diameter, the Stefan--Boltzmann relation gives
$T_{\rm eff,SB}\simeq2900$~K. This is slightly below the commonly adopted
$\sim3000$~K scale for Proxima but higher than our
$T_{\rm eff,SB}=2833\pm49$~K result.

The later empirical relation between absolute $K_s$-band magnitude, $M_{K_s}$, and mass from \citet{Mann2019} agrees with the earlier \citet{Mann2015} calibration to within about 5\% over their common mass range. Proxima's estimated mass, $M_\star=0.1221\pm0.0022\,M_\odot$, places it near the two lowest-mass companions in the 23-object eclipsing-binary sample of \citet{Swayne2024}. These stars have masses of $0.120$ and $0.129\,M_\odot$ and effective temperatures of $2699\pm59$ and $2822\pm83$~K, respectively. Our $T_{\rm eff,SB}=2833\pm49$~K agrees closely with the hotter of the two, showing that such a temperature is empirically plausible at Proxima's mass.

Our PHOENIX limb-darkened diameter has a fractional uncertainty of 2.3\%. Future VLTI observations with BIFROST \citep{Kraus2022} could reduce this uncertainty by resolving Proxima more strongly and constraining its limb darkening more directly.

\section{Conclusions}\label{sec:conclusions}

We analyzed five VLTI/PIONIER epochs: two from 2014 that were previously published by \citet{Rabus2019} and \citet{Lachaume2019}, and three previously unpublished epochs from 2017 and 2019.  The joint uniform-disk fit gives $\theta_{\rm UD}=1.117\pm0.024$~mas, and the PHOENIX Stefan--Boltzmann-consistent limb-darkened fit gives $\theta_{\rm LD,PHX}=1.154\pm0.026$~mas. After correcting the Gaia EDR3 parallax, $\varpi=768.089\pm0.050$~mas, we use $\theta_{\rm LD,PHX}$ with the Gaia distance to obtain $R_\star=0.162\pm0.004\,R_\odot$, and with the adopted bolometric flux to obtain $T_{\rm eff,SB}=2833\pm49$~K.
The joint uniform-disk diameter is $4.8\%$ larger than the 2014 value reported by \citet{Rabus2019} ($2.0\sigma$). The companion search finds no significant additional $H$-band source within the searched angular region and contrast range.
More broadly, Proxima remains an important low-mass benchmark: future interferometric observations, including with BIFROST \citep{Kraus2022}, together with improved spectroscopic and evolutionary models, will provide increasingly stringent tests of M-dwarf atmospheres, radii, and interiors.

\begin{acknowledgments}
We thank the reviewer for their careful reading and constructive suggestions, which helped us improve the clarity of the data-fitting methodology.
We thank Russel White for helpful discussions that improved the presentation of this paper. This work is based on archival observations collected at the European Organisation for Astronomical Research in the Southern Hemisphere under ESO programmes 092.D-0647, 093.D-0471, 299.C-5039(B), and 0102.D-0562(A). We thank the developers and maintainers of PIONIER and the \texttt{pndrs} reduction software. The data analysis and visualizations were carried out in \texttt{Python} using the open-source packages
\texttt{NumPy} \citep{Harris2020}, \texttt{SciPy} \citep{Virtanen2020},
\texttt{pandas} \citep{McKinney2010}, \texttt{Matplotlib} \citep{Hunter2007},
and \texttt{Astropy} \citep{Astropy2013,Astropy2018,Astropy2022}.
NA acknowledges the National Science Foundation under Grant Nos. AST-2034336 and AST-2407956.
AG acknowledges support from the Agencia Nacional de Investigaci\'on y Desarrollo (ANID) through FONDECYT Regular grant 1241073.
ChatGPT (v5.5) was used for language editing \footnote{\url{https://openai.com/chatgpt/}}.
\end{acknowledgments}

\facilities{VLTI:PIONIER}
\software{\texttt{pndrs (v3.94)}; \texttt{PMOIRED} (v1.3.16); \texttt{ExoTiC-LD (v3.2.1; data v3.2.0)}; \texttt{CANDID}; \texttt{astroARIADNE}; \texttt{NumPy}; \texttt{SciPy}; \texttt{pandas}; \texttt{Matplotlib}; \texttt{Astropy}}

\appendix
\section{Transfer-Function Calibration}\label{sec:transfer-function}
\begin{figure*}[!ht]
\centering
\includegraphics[width=0.66\textwidth]{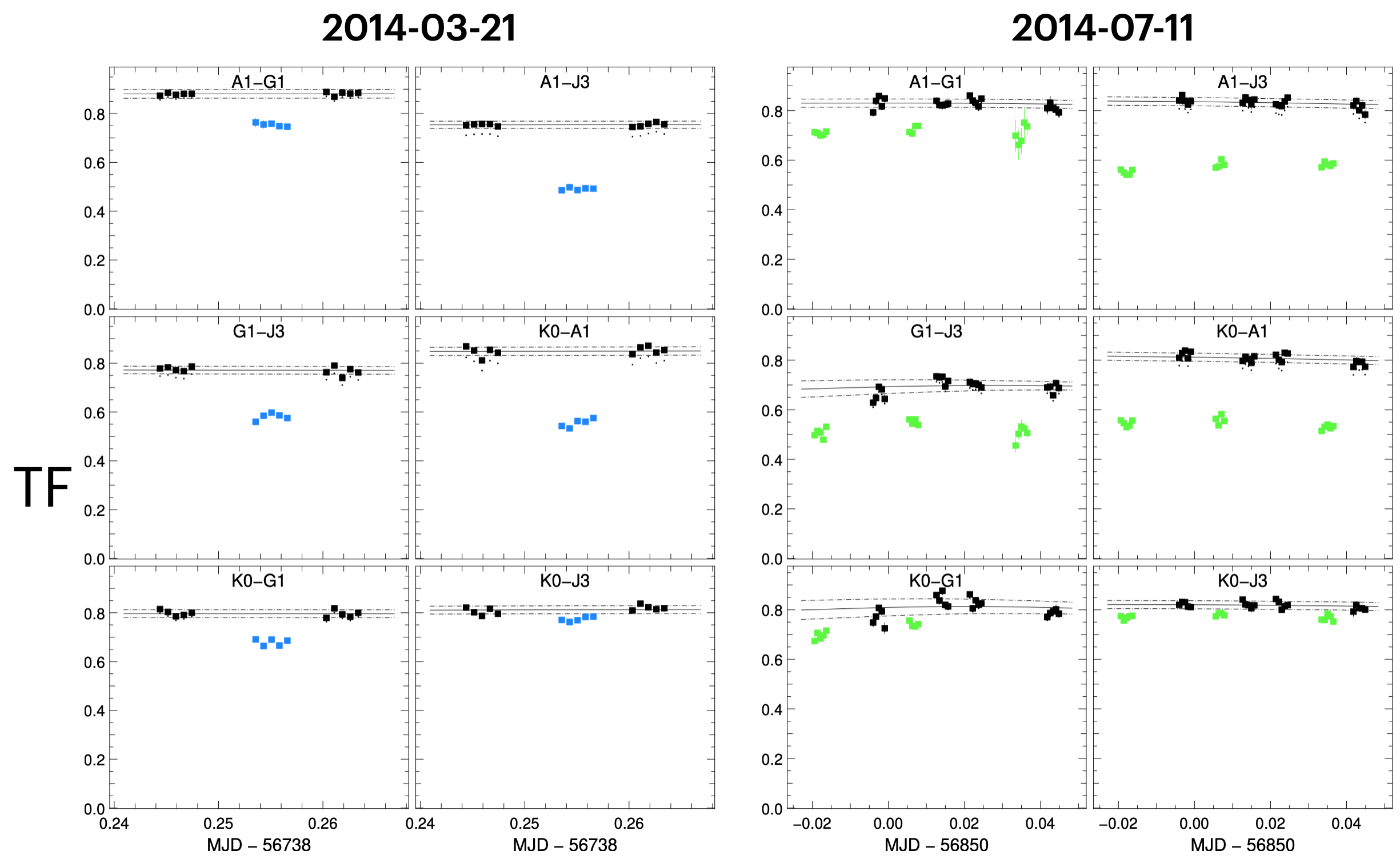}
\includegraphics[width=\textwidth]{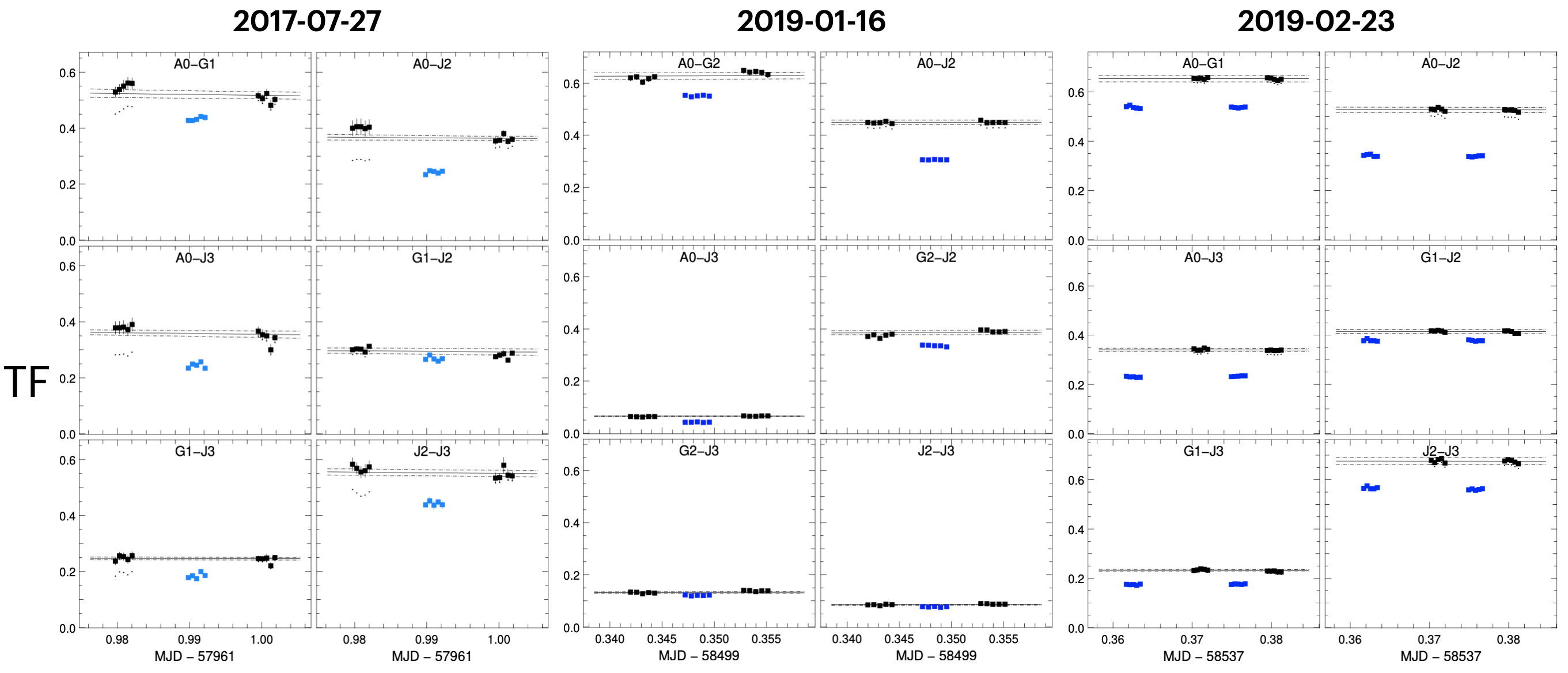}
\caption{Squared-visibility transfer functions from the reductions of the five epochs listed in Table~\ref{tab:calibrators}. The upper panel shows the two 2014 epochs and the lower panel shows the 2017 and 2019 epochs. The ordinate is dimensionless $V^2$: for the black calibrator points it represents the transfer function, $\mathcal{T}_{V^2}=V^2_{\rm cal,meas}/V^2_{\rm cal,model}$, whereas the blue points are raw Proxima science $V^2$ measurements before division by the interpolated transfer function.  The gray horizontal lines mark the transfer-function reference level and its bracketing values; the solid line is the central reference, while the dashed lines show the upper and lower bounds.}
\label{fig:transfer-function}
\end{figure*}

Figure~\ref{fig:transfer-function} shows the transfer function.
The transfer function describes the wavelength- and baseline-dependent combination of instrumental contrast loss and atmospheric visibility loss. Following the standard PIONIER calibration procedure, for example,  described by \citet{Rains2020}, a calibrator with an adopted angular diameter is used to predict its expected visibility, $V_{\rm cal,model}(B,\lambda)$. The measured calibrator visibility then defines
\begin{equation}
 {\cal T}(t,B,\lambda)=
 \frac{V_{\rm cal,meas}(t,B,\lambda)}
 {V_{\rm cal,model}(B,\lambda)} ,
\end{equation}
where $\mathcal{T}$ is the visibility-amplitude transfer function. 

The calibrated science visibility is obtained by dividing the measured target visibility by the transfer function interpolated in time between the bracketing calibrator observations,
\begin{equation}
V_{\rm sci,cal}(t,B,\lambda)=
\frac{V_{\rm sci,meas}(t,B,\lambda)}{{\cal T}(t,B,\lambda)} 
\end{equation}

For the squared-visibility, the corresponding transfer function is $\mathcal{T}_{V^2}=\mathcal{T}^2=V^2_{\rm cal,meas}/V^2_{\rm cal,model}$, and the calibrated squared visibility used in the diameter fits is $V^2_{\rm sci,cal}=V^2_{\rm sci,meas}/\mathcal{T}_{V^2}$. The calibrator-diameter uncertainties and the transfer-function interpolation uncertainty are propagated by \texttt{pndrs}.


In Figure~\ref{fig:transfer-function}, the two 2014 epochs are in the upper panel, and the three epochs from 2017 and 2019 are shown in the lower panel.

\pagebreak
\bibliography{references}
\bibliographystyle{aasjournal}
\end{document}